\def\draftversion{true}

\RequirePackage{ifthen}
\ifthenelse{\equal{\draftversion}{false}}{
  \documentclass[galley,aps,pra,10pt,amsmath,amssymb,
    superscriptaddress,nofootinbib,longbibliography,floatfix]{revtex4-2}
}{
  \documentclass[twocolumn,aps,prb,10pt,amsmath,amssymb,
    superscriptaddress,nofootinbib,longbibliography,floatfix]{revtex4-2}
}

\usepackage{graphicx}
\usepackage[usenames,dvipsnames]{color} 
\usepackage{bm} 
\usepackage{bbm} 
\usepackage{soul} 
\usepackage{mathtools}
\usepackage[makeroom]{cancel}
\usepackage{xspace}
\usepackage{multirow}
\usepackage{setspace}
\usepackage{hyperref}
\hypersetup{%
    colorlinks=true,       
    linkcolor=blue,
    citecolor=blue,       
    filecolor=magenta,      
    urlcolor=blue          
}

\ifthenelse{\equal{\draftversion}{true}}{
  \marginparwidth 2.7in
  \marginparsep 0.5in
  \newcounter{comm} 
  \def\commnext{\stepcounter{comm}}
  \def\commtext{{\bf\color{blue}[\arabic{comm}]}}
  \def\commmar{{\bf\color{blue}[\arabic{comm}]}}
  \def\aum#1{\commnext\marginpar{\small AU\commmar: #1}\commtext}
  \def\dsm#1{\commnext\marginpar{\small DS\commmar: #1}\commtext}
  \def\ytm#1{\commnext\marginpar{\small YT\commmar: #1}\commtext}
  \def\spm#1{\commnext\marginpar{\small SP\commmar: #1}\commtext}
  \def\srm#1{\commnext\marginpar{\small SR\commmar: #1}\commtext}
  \def\krm#1{\commnext\marginpar{\small KR\commmar: #1}\commtext}

}{
  \def\aum#1{}
  \def\dsm#1{}
  \def\ytm#1{}
  \def\spm#1{}
  \def\srm#1{}
  \def\krm#1{}

}

\newcommand{\beq}{\begin{equation}}
\newcommand{\eeq}{\end{equation}}
\newcommand{\bea}{\begin{eqnarray}}
\newcommand{\eea}{\end{eqnarray}}

\newcommand{\code}[1]{\textsc{#1}}   

\begin{document}

\title {Systematic display of the spin splitting in band structures of representative altermagnetic crystals }

\author{Mesfin Eshete}
\affiliation{Department of Physics and Astronomy, Center for Materials Theory, Rutgers University, 
Piscataway, New Jersey 08854, USA}%
\affiliation{Department of Industrial Chemistry, Addis Ababa Science and Technology University, P.O.Box 16417, Addis Ababa, Ethiopia}

\author{Yujia Teng}
\affiliation{Department of Physics and Astronomy, Center for Materials Theory, Rutgers University, 
Piscataway, New Jersey 08854, USA}%

\author{Andrea Urru}
\affiliation{Department of Physics and Astronomy, Center for Materials Theory, Rutgers University, 
Piscataway, New Jersey 08854, USA}%
\affiliation{Dipartimento di Fisica, Università di Cagliari, Cittadella Universitaria, Monserrato, CA 09042, Italy}

\author{Daniel Seleznev}
\affiliation{Department of Physics and Astronomy, Center for Materials Theory, Rutgers University, 
Piscataway, New Jersey 08854, USA}%
\affiliation{Department of Physics, University of Texas at Austin, Austin, Texas 78712, USA}

\author{Se Young Park}
\affiliation{Department of Physics and Origin of Matter and Evolution of Galaxies (OMEG) Institute, Soongsil University, Seoul 06978,  Korea}

\author{Sebastian E. Reyes-Lillo}
\affiliation{Departamento de F\'isica y Astronom\'ia, Universidad Andres Bello, Santiago 837-0136, Chile}

\author{Karin M. Rabe}
\email{kmrabe@physics.rutgers.edu}
\affiliation{Department of Physics and Astronomy, Center for Materials Theory, Rutgers University,
Piscataway, New Jersey 08854, USA}

\begin{abstract}
In this work we demonstrate a novel approach to exhibit the unique spin splitting that is typical of collinear altermagnets. This approach is to plot band structures on Brillouin zone paths that sample general k-points to show a representative picture that corresponds to Brillouin zone averages. This is in contrast to conventional band structure plotting which plots band structures on the highest-symmetry points and lines, and thus in many cases can fail to show any altermagnetic spin splitting at all. Our investigation compares the new approach with the band structures of collinear altermagnets  using conventional band structure plotting. We report the band structure and symmetry analysis for MnTe, CrSb, SmFeO$_3$, ScCrO$_3$, LaMnO$_3$, TlCrO$_3$, HoFeO$_3$, InCrO$_3$ and DyFeO$_3$. This result clearly demonstrates the advantage of this novel method for displaying the spin splitting of collinear altermagnets.

\end{abstract}

\maketitle
Renewed interest in unconventional magnetic ordering in crystals~\cite{chen2026rise,vsmejkal2022beyond} has been sparked by the recent discovery of altermagnetism, a previously unrecognized type of antiferromagnetism \cite{vsmejkal2022giant,fender2025altermagnetism, song2025altermagnets}. In all antiferromagnets, symmetry enforces zero macroscopic magnetization.
In conventional antiferromagnets, symmetry further enforces double degeneracy of the electronic states at all wavevectors in the Brillouin zone (BZ). In altermagnets (AMs), generally assumed to be collinear~\cite{song2025altermagnets,fender2025altermagnetism,jungwirth2016antiferromagnetic,cheong2025altermagnetism,vsmejkal2022emerging}, symmetry enforces double degeneracy only for certain high-symmetry wavevectors. Because electronic states are spin split even when magnetization is zero, AMs are considered highly promising for spintronic applications, potentially enabling robust charge-spin conversion, fast switching, greater insensitivity to destabilizing fields and lower energy usage for overall enhanced performance \cite{ jungwirth2016antiferromagnetic,papaj2023andreev,watanabe2024symmetry,chakraborty2024zero,noh2025tunneling,leiviska2025spin,samanta2025spin,sourounis2025efficient,guo2025mechanically,Chen_2024,baral2023giant}. 

The symmetry of a crystal with collinear magnetic order is fully defined by its ``magnetic space group (MSG) without spin-orbit coupling (SOC)'' \cite{yuan2020giant,yuan2021prediction,yuan2023uncovering,turek2022altermagnetism,yuan2024nonrelativistic}. The symmetry criterion for a collinear antiferromagnet to be altermagnetic is readily established by considering the symmetry operations that enforce zero magnetization. 
If $U\boldsymbol t$ or $PT$ or both are symmetries, where $U$ is a 180$^\circ$ spin rotation, $\boldsymbol t$ is a translation, $P$ is space inversion and $T$ is the time-reversal operator, then the electronic energy levels at general $\boldsymbol k$  points are doubly degenerate and the system is a conventional antiferromagnet. If neither $U\boldsymbol t$ nor $PT$ is a symmetry, then the symmetry that enforces zero magnetization by interchanging up and down sublattices has spatial part $\{R\vert \boldsymbol t\}$, with $R$ neither the identity nor inversion; we refer to such $R$ as a ``spin-flip operation''. 
This symmetry results in the levels at $\boldsymbol k'=R \boldsymbol k$ being the spin-flipped versions of the levels at $\boldsymbol k$. 

The fundamental reason why AM was not previously recognized as a distinct type of magnetic order is that AMs generally have symmetry-enforced zero spin splitting on high-symmetry lines and planes in the BZ. As a result, it is expected that spin splitting will not be visible in typical band structure plots,  which trace a path around the BZ generated by high-symmetry lines. 

 We argue that for AMs, a different approach to plotting band structures is needed. In particular, it is essential to sample interior points of the irreducible BZ; such sampling can have the additional advantage of making the BZ averages apparent at a glance. Here, we do this using a path in reciprocal space constructed by a method we call AlterSeeK-Path~\cite{teng2026,alterseek-path}, which incorporates general lines into a conventional path of high-symmetry lines~\cite{seek-path}. Specifically, we choose a general k-point ($\boldsymbol k$) as the volume centroid of the irreducible Brillouin zone (IBZ) and then apply a spin-flip operation $R$ to generate $\boldsymbol k'$. We then introduce $\boldsymbol k$ and $\boldsymbol k'$ into the conventional high-symmetry path to display the alternating nature of the collinear altermagnetic spin splitting. As shown previously for altermagnetic BiFeO$_3$~\cite{urru2025g}, this results in a band structure plot that clearly demonstrates the presence of spin splitting and sign alternation across the BZ.

In this paper, we present band structures for a set of collinear altermagnetic crystals plotted using our systematic approach and compare them to conventional band structure plots. From this comparison, we conclude that the systematic approach is more useful in understanding the size and character of altermagnetic spin splitting and in estimating physical properties based on averages over the BZ, particularly those associated with altermagnetism.

\begin{figure} 
    \centering
    \includegraphics[width=0.45\textwidth]{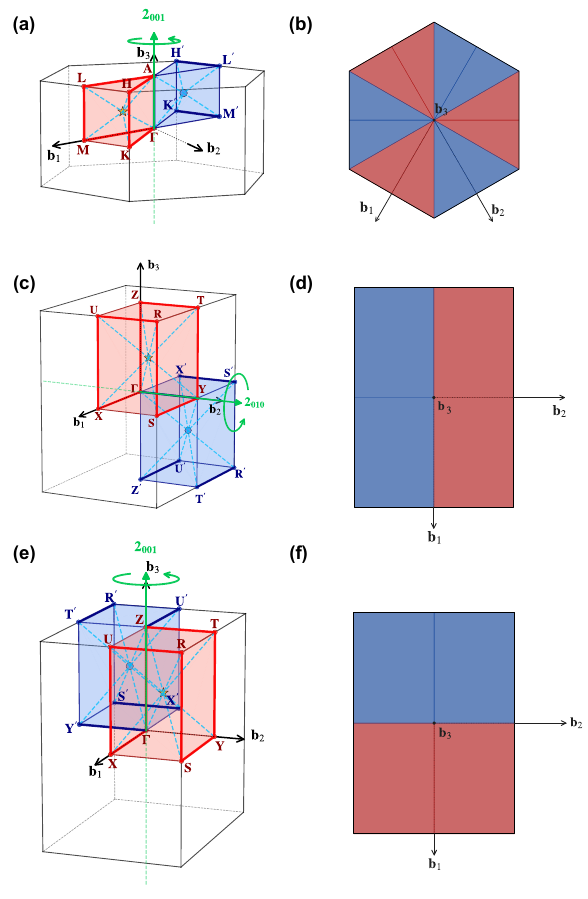} 
    \caption{(a), (c) and (e) BZ with the irreducible BZ in red and its image under the spin-flip operation shown in blue, for the hexagonal and orthorhombic compounds, respectively. The top view of the altermagnetic spin pattern is shown in (b) for the hexagonal compounds (bulk g-wave character), and in (d) for the orthorhombic SmFeO$_3$ (bulk d-wave character) (f) for the orthorhombic ScCrO$_3$ (bulk d-wave character).}
    \label{fig:BZ}
\end{figure}

 \begin{figure*} 
    \centering
    \includegraphics[width=0.8\textwidth]{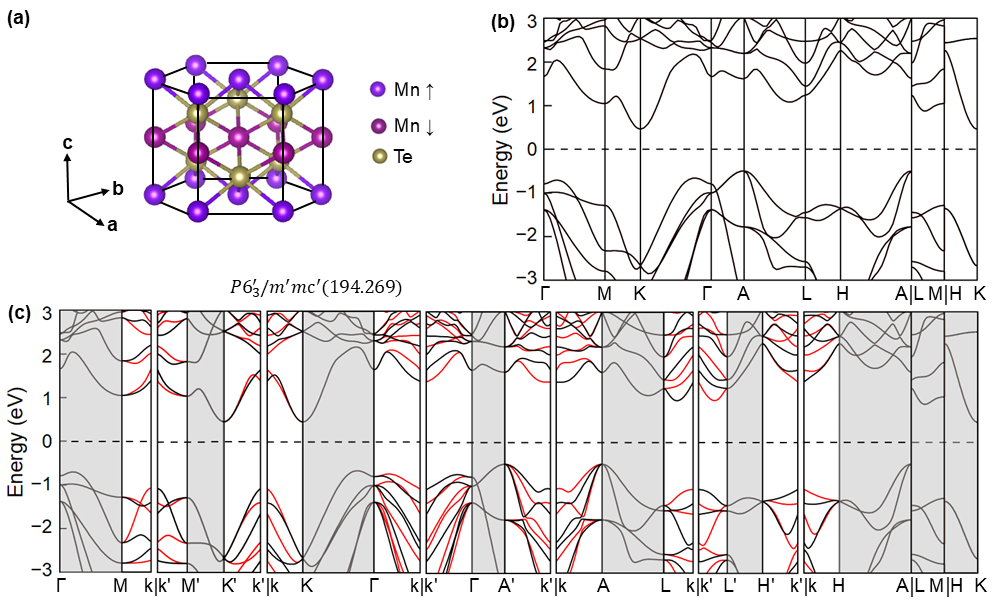} 
    \caption{(a) The hexagonal crystal structure of MnTe with its magnetic information. The band structures of MnTe along high-symmetry lines (b) with DFT+U (U = 6~eV) computation. The band structures of MnTe calculated without SOC along a generalized path in reciprocal space, including general lines and high-symmetry lines (c) with DFT+U (U = 6~eV) calculation.}
    \label{fig:wide_figure2}
\end{figure*}

The set of collinear altermagnetic crystals chosen for this study --- MnTe, CrSb, SmFeO$_3$, ScCrO$_3$, LaMnO$_3$, TlCrO$_3$, HoFeO$_3$, InCrO$_3$ and DyFeO$_3$ --- is drawn from a recent high-throughput screening approach that combined embedded dynamical mean-field theory and density functional theory (DFT) to speed up the search for altermagnetic metals in the Bilbao MAGNDATA database \cite{k47t-23gp,gallego2016magndata}. The crystals used for this study are semiconducting/insulating (MnTe, SmFeO$_3$, ScCrO$_3$, LaMnO$_3$, TlCrO$_3$, HoFeO$_3$, InCrO$_3$, and DyFeO$_3$) and metallic (CrSb). They also have different lattice types (hexagonal, orthorhombic) that enable comparisons of symmetry effects. 
Crucially, these crystals are accessible both theoretically and experimentally.  

For each compound, we generate the first-principles band structure according to the following steps. First, we obtain the atomic positions from the mcif file in the Bilbao MAGNDATA database \cite{gallego2016magndata}. Next, we use the \code{FINDSPINGROUP} package \cite{yu2026identifyingorientedspinspace} to analyze the magnetic ordering information to confirm the MSG and that this magnetically ordered crystal is altermagnetic \cite{PhysRevX.14.031038}. From this analysis, we also identify a particular spin-flip operation, which is a symmetry operation that combines a spin-flip with an interchange of the up sublattice of the magnetic atoms and the down sublattice. Then, we use our package AlterSeeK-Path \cite{teng2026} to identify the ``MSG without SOC'' and k-point path in the BZ for systematic visualization of the spin splitting in the band structure. As described in the introduction, this path combines the conventional high-symmetry lines with lines that connect a general k-point $\boldsymbol k$, chosen as the centroid of the IBZ, and the high-symmetry points on the vertices of the IBZ, in conjunction with the images of this point ($\boldsymbol k'$) and these lines under the point part of the spin-flip operation. The two paths relevant to the hexagonal and orthorhombic compounds considered in this paper are shown in Fig.~\ref{fig:BZ}(a), (c) and (e), with the corresponding spin alternating patterns in Fig.~\ref{fig:BZ}(b), (d) and (f).

\begin{figure*} 
    \centering
    \includegraphics[width=0.8\textwidth]{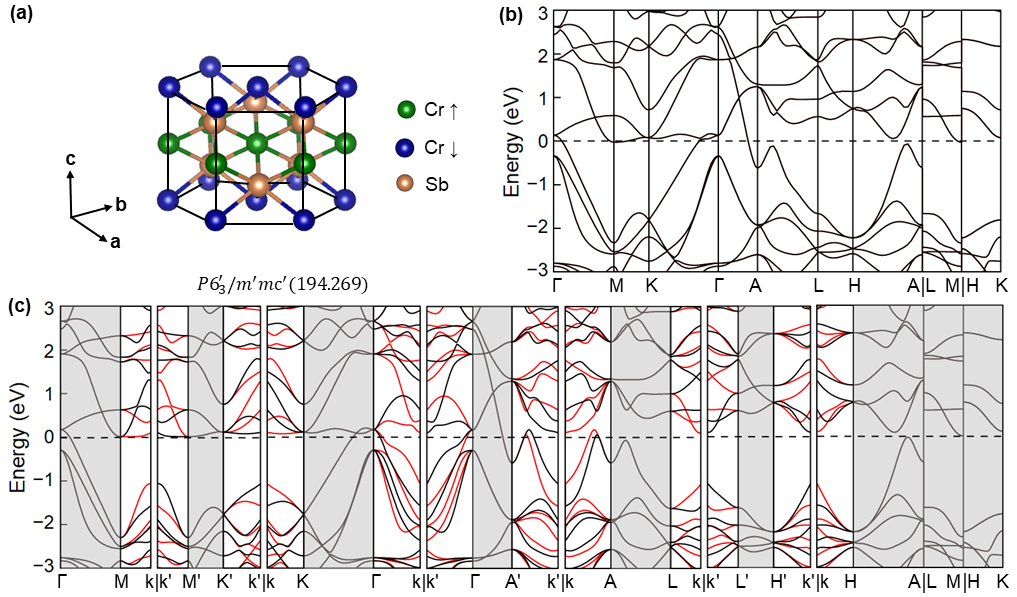} 
\caption{(a) The Hexagonal crystal structure of CrSb with its magnetic information.  The band structures of CrSb along high-symmetry lines (b) by DFT+U (U = 4~eV) computation. The band structures of CrSb calculated without SOC along a generalized path in reciprocal space, including general lines and high-symmetry lines (c) DFT+U (U = 4~eV) calculation.}
    \label{fig:wide_figure3}
\end{figure*}

Experimental work on MnTe shows it possesses a hexagonal structure with parent space group $P6_3/mmc$ (No.~194) and collinear A-type antiferromagnetic order with MSG $Cmcm$ (63.457) below the Néel temperature of 310 K~\cite{kunitomi1964neutron, g2025emergent,ferrer2000temperature,gallego2016magndata,k47t-23gp}.  This group does not include $PT$ or any operation of the form $U\boldsymbol t$, and thus this phase is altermagnetic. Previous first-principles investigations have established that this observed altermagnetic phase of MnTe is the ground state structure \cite{chernov2025electronic,osumi2024observation}.

In Fig.~\ref{fig:wide_figure2}(b), we present the first-principles DFT+U conventional band structure, plotted on high-symmetry lines, for MnTe with ``MSG without SOC'' is $P6'_3/m'mc'$ (194.269), which shows zero spin splitting. In previous first-principles studies of MnTe, altermagnetic spin splitting was shown by adding the line $\Gamma$-$L$ to the conventional band structure plot \cite{chernov2025electronic,osumi2024observation,k47t-23gp}.

In Fig.~\ref{fig:wide_figure2}(c), we present the band structure of MnTe, plotted using our new approach. The gray panels show the high-symmetry lines with zero spin splitting.  Added lines that connect high-symmetry points to the chosen general k-point $\boldsymbol k$ = (0.278, 0.111, 0.250), which is the centroid of the irreducible wedge (Fig.~\ref{fig:BZ}(a)), provide a representative sampling of the interior of the BZ. Spin splitting can clearly be seen on these interior lines. The splitting correctly changes sign in the image of each such line under the chosen spin-flip operation $R = 2_{001}$, which interchanges the up and down Mn sublattices. This pairing of lines demonstrates the spin splitting alternating across the BZ and the zero net magnetization. Our methodology thus yields clear information about the spin splitting in collinear altermagnetic MnTe. 

\begin{figure*} 
    \centering
    \includegraphics[width=0.8\textwidth]{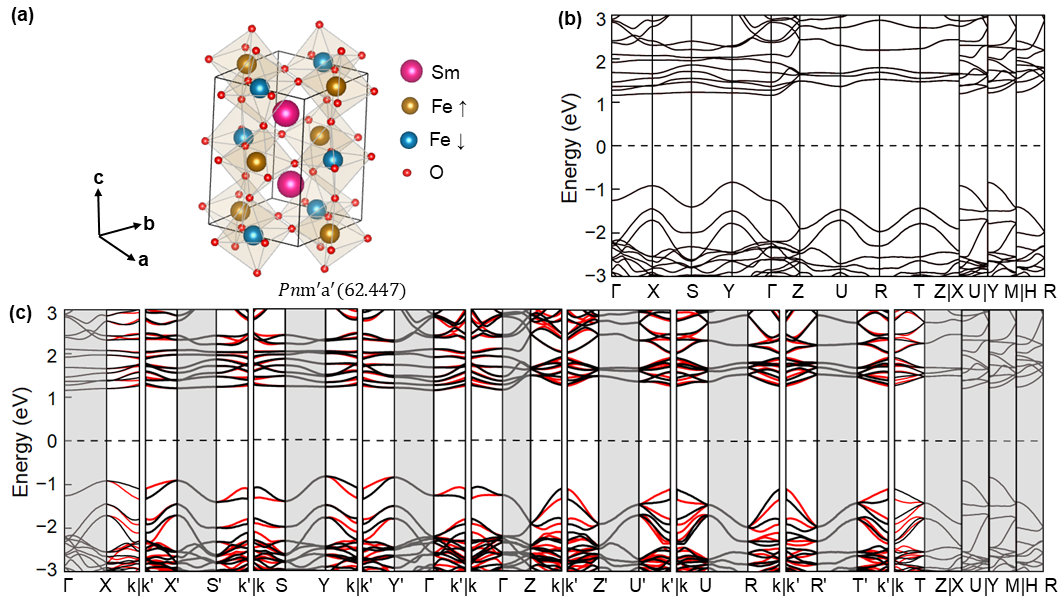} 
\caption{(a) The crystal structure of SmFeO$_3$ with its magnetic information. The band structures of SmFeO$_3$ along high-symmetry lines (b) DFT+U (U = 5.3~eV) computation. The band structures of SmFeO$_3$ calculated without SOC along a generalized path in reciprocal space, including general lines and high-symmetry lines (c) DFT+U (U = 5.3~eV) calculation.}
    \label{fig:wide_figure4}
\end{figure*}
We note that our DFT computation underestimates the band gap of the semiconducting MnTe, resulting in a metal, while the use of DFT+U improves the band gap significantly, as shown in Fig.~\ref{fig:wide_figure2}(b)-(c) and Table S1, respectively. Specifically, the incorporation of Hubbard U shifts the conduction band significantly and results in band gaps of 0.83 eV, 0.96 eV, and 1.05 eV with U values of 4, 6, and 8~eV, respectively. These results show that for MnTe, DFT+U is highly sensitive to the selection of U, consistent with previous works \cite{k47t-23gp}.

It is well known that CrSb exhibits antiferromagnetism at temperatures as high as 700 K~\cite{bean1962magnetic}. The magnetic ordering is reported as collinear A-type antiferromagnetic: the parent crystal has the hexagonal $P6_3/mmc$ (No.~194) space group, and the Cr spins in each layer are ordered ferromagnetically along the c direction, with the spin direction alternating from layer to layers, resulting in the observed MSG $P6'_3/m'm'c$ (194.268)~\cite{yuan2020magnetic, thadathil2026electrical,k47t-23gp,gallego2016magndata}.

The ``MSG without SOC'' for the reported magnetic ordering of  CrSb is $P6'_3/m'mc'$ (194.269). As expected from our discussion above for MnTe, which has the same ``MSG without SOC'', the conventional band structure plot for collinear altermagnetic CrSb (Fig.~\ref{fig:wide_figure3}(b)) shows no spin splitting. In a previous study, the altermagnetic spin splitting was shown by adding $\Gamma$-$L$ and $P$-$Q$-$P$ to the conventional path~\cite{k47t-23gp}, as for MnTe.

Our band structure, plotted using the AlterSeeK-Path construction~\cite{teng2026}, explicitly shows the altermagnetic spin splitting for the CrSb crystal, as illustrated in Fig.~\ref{fig:wide_figure3}(c). In contrast to MnTe, which is a semiconductor,  the results for CrSb show that the Fermi level intersects several bands, producing electron and hole pockets. From the intersections along interior BZ lines, we can identify parts of the Fermi surface with altermagnetic spin splitting, which makes CrSb a possible spin filter~\cite{terashima2026altermagnetic,reimers2024direct,li2025topological}.  In particular, we see large spin splitting along $\Gamma-k|k'-\Gamma$, and also note that there is a spin-split hole pocket along $A'-k'|k-A$.

Experimental works showed SmFeO$_3$ in the orthorhombic distorted perovskite structure with $Pnma$ (No.~62) space group. The structure features corner-sharing FeO$_6$ octahedra with strong tilting and rotation. The Fe spins exhibit collinear G-type antiferromagnetic (AFM) ordering at $T_\mathrm{N} \approx 670$~K with a reported easy axis rotation transition at $T_\mathrm{SR} \approx 480$~K \cite{kuo2014k, maslen1996synchrotron}. 
We note that in other findings, the observed magnetic ordering is noncollinear, with the spins slightly tilted away from the c axis and antiparallel along the a axis \cite{marshall2012magnetic,lee2011spin}.

The reported MSG for this collinear ordering is $Pn'ma'$ (62.448)~\cite{gallego2016magndata,kuo2014k,k47t-23gp}, with G-type AFM order. Our symmetry analysis shows that the ``MSG without SOC'' is  $Pnm'a'$ (62.447), as shown in Fig.~\ref{fig:wide_figure4}(a). This magnetically ordered orthorhombic structure is centrosymmetric but lacks $PT$ symmetry and $U\boldsymbol t$ symmetry, making it altermagnetic. $2_{010}$ can be chosen as the spin-flip operation $R$, interchanging the up and down Fe sublattices. The conventional band structure plot shows no spin splitting; in Ref.~\onlinecite{k47t-23gp}, the altermagnetic spin splitting was demonstrated by adding the lines $\Gamma$-$R$, $T$-$U$, and $X$-$Y$ to the conventional plot.
\begin{figure*} 
    \centering
    \includegraphics[width=0.8\textwidth]{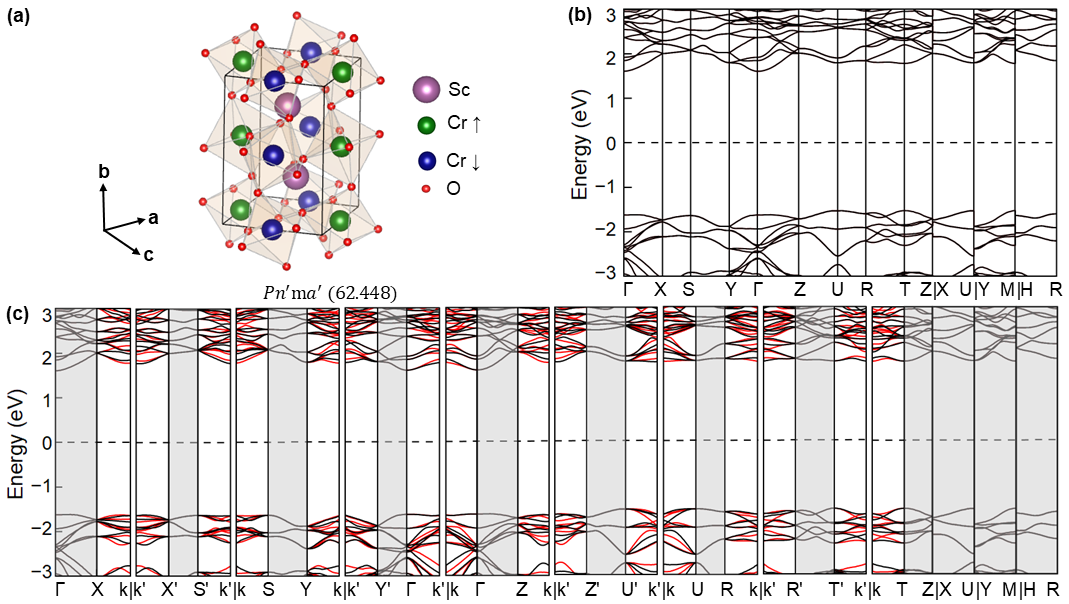} 
\caption{(a) The crystal structure of ScCrO$_3$ with its magnetic information. The band structures of ScCrO$_3$ along high-symmetry lines (b) DFT+U (U = 4~eV) computation. The band structures of ScCrO$_3$ calculated without SOC along a generalized path in reciprocal space, including general lines and high-symmetry lines (c) DFT+U (U = 4~eV) calculation.}
    \label{fig:wide_figure5}
\end{figure*}

The band structure plotted using our approach is presented in Fig.~\ref{fig:wide_figure4}(c).  $\boldsymbol k$ is selected with computed centroid inside the irreducible wedge with values $\boldsymbol k$ = (0.25, 0.25, 0.25) which is far from the high-symmetry planes and lines. Our band structure plot clearly shows spin splitting, especially in the top valence band. With a value for Hubbard U of 5.3 eV on Fe, the computed band gap is 2.0 eV as presented in Fig.~\ref{fig:wide_figure4}(b)-(c) and Table S1.

Experimental work on ScCrO$_3$ shows an orthorhombic structure with parent space group $Pnma$ (No.~62) and MSG $Pnma$ (62.441), having collinear C-type antiferromagnetic order below 70 K~\cite{ding2017unusual, gallego2016magndata,belik2012crystal,k47t-23gp}. We find that the ``MSG without SOC'' is $Pn'ma'$ (62.448), as shown in Fig.~\ref{fig:wide_figure5}(a). Like $Pnm'a'$ (62.447) discussed above, this group lacks $PT$ symmetry and $U\boldsymbol t$ symmetry, making it altermagnetic. The spin-flip operation $R = 2_{010}$ exchanges spin-up and spin-down Cr sublattices. The conventional band structure plot shows no spin splitting; in Ref.~\onlinecite{k47t-23gp}, the altermagnetic spin splitting was demonstrated by adding the lines $\Gamma$-$R$ and $S$-$R$ to the conventional plot. 

The band structure plotted using our approach is presented in Fig.~\ref{fig:wide_figure5}(c), where spin splitting can clearly be seen on the BZ interior lines and also the conventional band structure in the Fig.~\ref{fig:wide_figure5}(b) which shows no splitting. The inclusion of 4 eV Hubbard U on Cr raises the band gap from 1.38 eV (in the DFT calculation) to 3.13 eV.

We computed band structures using our approach for five additional orthorhombic compounds: LaMnO$_3$, TlCrO$_3$, HoFeO$_3$, InCrO$_3$, and DyFeO$_3$, identified as altermagnets in the high-throughput search in Ref.~\onlinecite{k47t-23gp}. The results are presented in Figs. S2-S6. 

The band structures presented in the main text and in the SM show that for altermagnets, conventional band structure plotting along high-symmetry lines fails: it generally shows no spin splitting at all. The ad hoc solutions described above are limited to confirming altermagnetism by showing nonzero spin splitting at some sufficiently low-symmetry $\boldsymbol k$ values.
Our new approach, by adding lines that systematically sample the BZ interior, not only confirms altermagnetism but also offers a representative sampling for BZ averages to identify and characterize hidden spin splitting for altermagnetic compounds.

While we have focused here on hexagonal and orthorhombic cases, which are relevant to altermagnets of particular current interest, these plots can be constructed for any altermagnetic crystal. These plots provide altermagnetic spin splitting and its variation inside the BZ at a glance. No other method to date has this capability. With the AlterSeeK-Path package~\cite{teng2026}, these systematic plots can be obtained just as easily as conventional band structure plots, while providing much more useful information about the altermagnetic spin splitting. In fact, we claim that our new approach, focusing on systematic sampling of the BZ interior, offers substantial advantages more generally, since physical properties involve BZ averages for which high-symmetry lines do not directly provide accurate information; this work is in progress.

In summary, we presented band structure plots for a representative set of altermagnetic crystals --- MnTe, CrSb, SmFeO$_3$, ScCrO$_3$, LaMnO$_3$, TlCrO$_3$, HoFeO$_3$, InCrO$_3$, and DyFeO$_3$ --- using a k-space path systematically constructed to display the characteristic spin splitting in the interior of the BZ. 
For all of these compounds, the conventional band structure path only includes high-symmetry lines on which spin splitting is zero, a problem which was previously addressed ad hoc by including additional band structure panel(s) along low-symmetry lines or by supplying an auxiliary plot showing a line in the interior of the BZ.  The calculations presented here demonstrate the benefit of using this new technique to show the spin splitting of collinear altermagnets, with negligible extra effort. We believe that this approach will improve the theoretical analysis of altermagnetic materials and aid in the search for new high-performance altermagnetic materials for practical applications.

\begin{acknowledgments}
M.E., A.U., D.S., and K.M.R. acknowledge support from the Office of Naval Research Grant No. N00014-21-1-2107. M.E. was also supported by the Gordon and Betty Moore Foundation. Y.T. acknowledges support from the Wisconsin Materials Research Science and Engineering Center (NSF DMR-2309000). A.U. was also supported by the Italian Fund for Science (FIS), grant No. FIS-2024-04793. D.S. was also supported by the W. M. Keck Foundation under grant 996588. S.Y.P. was supported by the National Research Foundation of Korea (NRF) grant funded by the Korea government (MSIT) (RS-2024-00358551). S.E.R.-L. acknowledges ANID Fondecyt regular grant number 1260824. Computational facilities were provided by the Beowulf cluster at the Department of Physics and Astronomy of Rutgers University.
\end{acknowledgments}

\appendix 
\section{Computational details}
First-principles density-functional-theory computations are carried out using the Vienna \textit{ab initio} simulation package (VASP)~\cite{vasp1,vasp2}, with the projector augmented wave (PAW)~\cite{paw} method and the Perdew--Burke--Ernzerhof (PBE) generalized gradient approximation (GGA) exchange-correlation functional~\cite{gga}.  A Hubbard U correction in the Dudarev formulation~\cite{dft+u} was applied as an effective U on the $d$ states of the magnetic transition-metal atoms, with U = 4, 6, and 8~eV for MnTe (Mn) and CrSb (Cr); U = 6~eV for LaMnO$_3$ (Mn); U = 4~eV for ScCrO$_3$, TlCrO$_3$, and InCrO$_3$ (Cr); and U = 5.3~eV for SmFeO$_3$, HoFeO$_3$, and DyFeO$_3$ (Fe), as summarized in Table S1. These values are taken from the high-throughput screening study of Ref.~\cite{k47t-23gp}, from which the present set of compounds is drawn. The energy cutoff was set to 520 eV and Brillouin zone integrations are performed on an $11 \times 11 \times 11$ $\Gamma$-centered Monkhorst--Pack mesh~\cite{k-points}. Self-consistent ground-state calculations are performed with an energy convergence threshold of $10^{-6}$~eV. All calculations are spin-polarized with collinear magnetic order and do not include SOC.
   
\bibliography{cite}

\end{document}



\title{Supplementary Material for ``Systematic display of the spin splitting in band structures of representative altermagnetic crystals''}

\author{Mesfin Eshete}
\affiliation{Department of Physics and Astronomy, Center for Materials Theory, Rutgers University,
Piscataway, New Jersey 08854, USA}%
\affiliation{Department of Industrial Chemistry, Addis Ababa Science and Technology University, P.O.Box 16417, Addis Ababa, Ethiopia}

\author{Yujia Teng}
\affiliation{Department of Physics and Astronomy, Center for Materials Theory, Rutgers University,
Piscataway, New Jersey 08854, USA}%

\author{Andrea Urru}
\affiliation{Department of Physics and Astronomy, Center for Materials Theory, Rutgers University, 
Piscataway, New Jersey 08854, USA}%
\affiliation{Dipartimento di Fisica, Università di Cagliari, Cittadella Universitaria, Monserrato, CA 09042, Italy}

\author{Daniel Seleznev}
\affiliation{Department of Physics and Astronomy, Center for Materials Theory, Rutgers University, 
Piscataway, New Jersey 08854, USA}%
\affiliation{Department of Physics, University of Texas at Austin, Austin, Texas 78712, USA}
\author{Se Young Park}
\affiliation{Department of Physics and Origin of Matter and Evolution of Galaxies (OMEG) Institute, Soongsil University, Seoul 06978,  Korea}

\author{Sebastian E. Reyes-Lillo}
\affiliation{Departamento de F\'isica y Astronom\'ia, Universidad Andres Bello, Santiago 837-0136, Chile}

\author{Karin M. Rabe}
\email{kmrabe@physics.rutgers.edu}
\affiliation{Department of Physics and Astronomy, Center for Materials Theory, Rutgers University,
Piscataway, New Jersey 08854, USA}

\maketitle
\section{Results for additional compounds}
In Table~\ref{tab:tab1} we present a list of all compounds studied. The first four, MnTe, CrSb, SmFeO$_3$ and ScCrO$_3$, were discussed in the main text.

\begin{table}[h!]
\caption{Table S1. The name of each material, parent space group (PSG) (with ID in parentheses), MSG (with ID in parentheses), MSG without SOC (with ID in parentheses), spin-flip operation R, bandgap  with DFT+U (in eV), Hubbard U value (in eV), and bandgap with DFT (in eV).}
    \begin{tabular}{|c|c|c|c|c|c|c|c|} 
    \hline
    Material &PSG &MSG & MSG without SOC & R & {bandgap (DFT+U)} & U & bandgap (DFT)\\
     \hline
     & & & & & 0.83 & 4 & Metallic\\ 
    MnTe &$P6_3/mmc$(194) & $Cmcm$(63.457)\cite{kunitomi1964neutron,k47t-23gp} & $P6'_3/m'mc'$(194.269)& $2_{001}$ & 0.96 & 6 & Metallic\\
     & & & & & 1.05 & 8 & Metallic\\
     \hline
     & & & & &Metallic & 4 & Metallic\\
    CrSb &$P6_3/mmc$(194) & $P6'_3/m'm'c$ (194.268)\cite{yuan2020magnetic,k47t-23gp}& $P6'_3/m'mc'$(194.269)&$2_{001}$ & Metallic & 6 & Metallic\\
     & &  & & & Metallic & 8 & Metallic\\
     \hline
    SmFeO$_3$ &$Pnma$(62) & $Pn'ma'$(62.448)\cite{kuo2014k,k47t-23gp} & $Pnm'a'$(62.447)&$2_{010}$ & 2.0 & 5.3 &  Metallic\\
     \hline
    ScCrO$_3$ &$Pnma$(62) &  $Pnma$(62.441)\cite{ding2017unusual,k47t-23gp} & $Pn'ma'$(62.448)&$2_{010}$ & 3.13 & 4 & 1.38\\
     \hline
    LaMnO$_3$ &$Pnma$(62) & $Pn'ma'$(62.448)\cite{moussa1996spin,k47t-23gp}& $Pn'm'a$(62.446) &$2_{010}$ & 1.54 & 6 & 0.18\\
    \hline
    TlCrO$_3$ & $Pnma$(62) & $Pnma$ (62.441)\cite{ding2017unusual,k47t-23gp}  & $Pn'ma'$(62.448) & $m_{100}$ & 0.52 & 4 & Metallic\\
     \hline
    HoFeO$_3$ &$Pnma$(62) & $Pn'ma'$(62.448)\cite{chatterji2017temperature,k47t-23gp} & $Pn'ma'$(62.448) &$2_{100}$ &0.86 & 5.3 & 0.19\\
     \hline
   InCrO$_3$ & $Pnma$(62)&$Pnma$(62.441)\cite{ding2017unusual,k47t-23gp} &$Pn'ma'$(62.448) &$m_{100}$ &1.73 & 4 & 0.94\\
   \hline
   DyFeO$_3$ & $Pnma$(62) & $Pnma$(62.441)\cite{ritter2022magnetic,k47t-23gp} &$Pn'ma'$(62.448) & $2_{100}$ & Metallic & 5.3 & Metallic \\
     \hline
     \end{tabular}
     \label{tab:tab1}
\end{table}
\begin{figure*}[t] 
    \centering
    \includegraphics[width=0.5\textwidth]{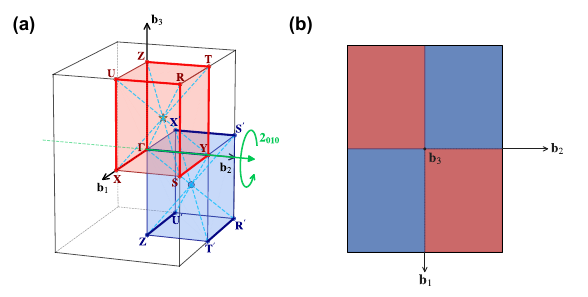}
\caption{(a) Brillouin zone (BZ) with the nonmagnetic irreducible BZ in red and its image under the spin-flip operation shown in blue, for LaMnO$_3$; (b) altermagnetic (bulk d-wave) pattern for LaMnO$_3$.}
\label{fig:figureS1}
\end{figure*}

\begin{justify}
LaMnO$_3$ is reported in an orthorhombic distorted perovskite structure with A-type antiferromagnetic order 
~\cite{moussa1996spin,norby1995crystal,moreno2008preparation}. 
This phase of LaMnO$_3$ has MSG $Pn'ma'$ (62.448)~\cite{k47t-23gp,gallego2016magndata,moussa1996spin}. From our symmetry analysis, this corresponds to MSG without SOC $Pn'm'a$ (62.446). The altermagnetic spin pattern is bulk d-wave, as shown in Fig.~\ref{fig:figureS1}(b) with the BZ shown in Fig.~\ref{fig:figureS1}(a).
As expected, the conventional plotting along high-symmetry lines shows zero splitting, depicted in Fig.~\ref{fig:wide_figureS2}(a). In Ref.~\onlinecite{k47t-23gp}, the altermagnetic spin splitting was demonstrated by adding the lines $\Gamma$-$R$, $X$-$Y$, and $Y$-$Z$, to the conventional plot.
Our systematic approach for band structure plotting shows spin splitting throughout the interior of the BZ, as seen in Fig.~\ref{fig:wide_figureS2}(b). We note that the strong correlations in LaMnO$_3$ require computation with DFT+U with a U value of 6 eV; in particular, this increases the DFT gap of 0.18 eV to 1.54 eV. 
\begin{figure*}[h] 
    \centering
    \includegraphics[width=0.8\textwidth]{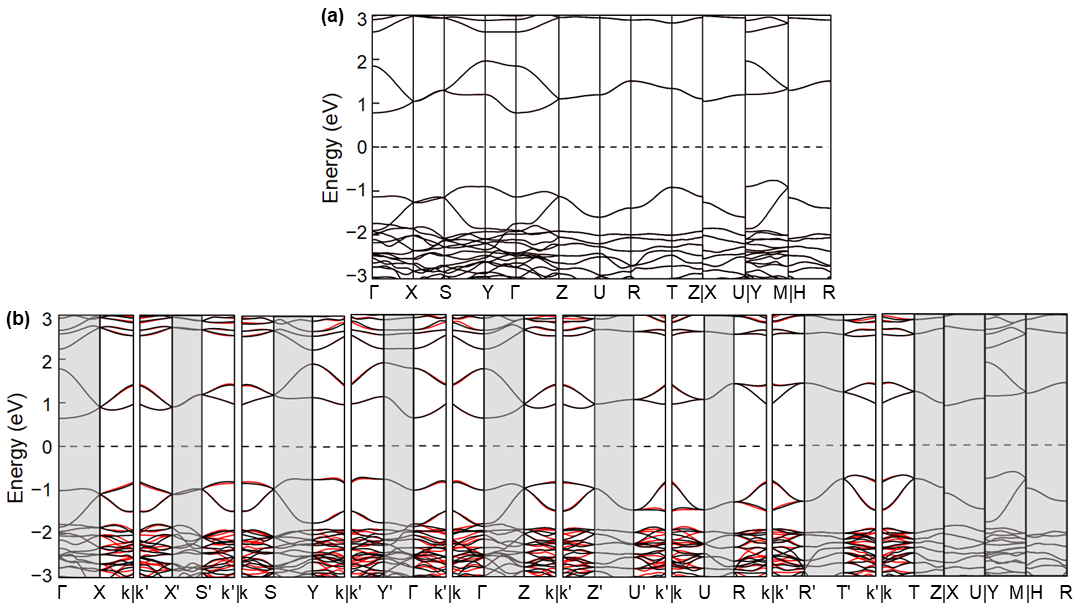} 
\caption{Nonrelativistic DFT+U band structures of orthorhombic LaMnO$_3$ (U = 6~eV) (a) The band structure along high-symmetry lines; (b) the band structure along the generalized BZ path constructed using AlterSeeK-Path~\cite{teng2026}.}
    \label{fig:wide_figureS2}
\end{figure*} 
\begin{figure*}[h]
    \centering
    \includegraphics[width=0.8\textwidth]{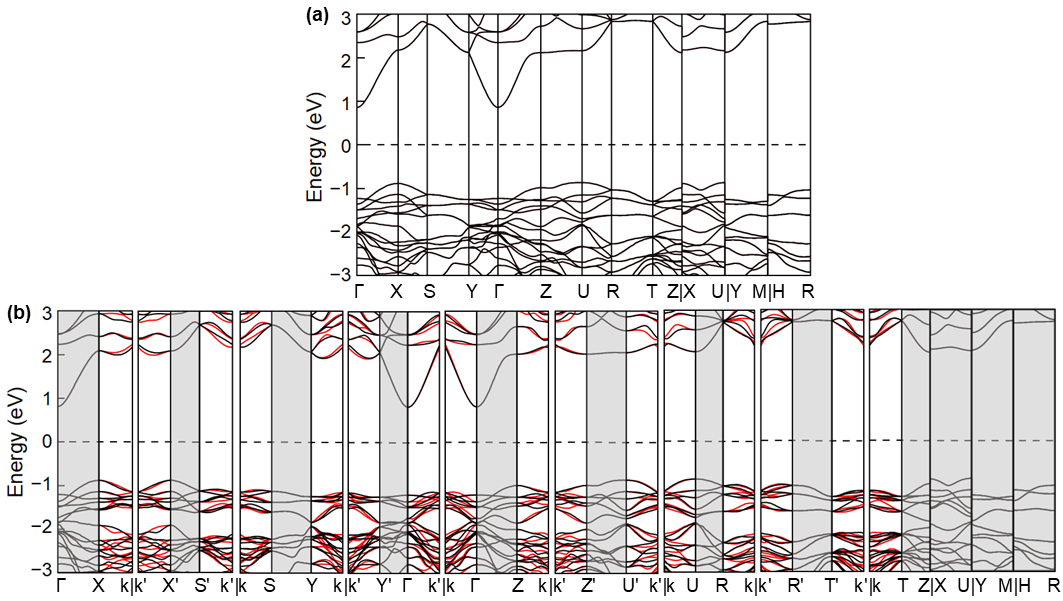} 
    \vspace{-0.5cm}
    \begin{justify}
     \caption {Nonrelativistic DFT+U band structures of orthorhombic TlCrO$_3$ (U = 4~eV) (a) The band structure along high-symmetry lines; (b) the band structure along the generalized BZ path constructed using AlterSeeK-Path.}   
    \label{fig:wide_figureS3}
    \end{justify}
\end{figure*}

Experimental and theoretical studies of TlCrO$_3$ show that it has an orthorhombic distorted perovskite structure (parent space group $Pnma$ (62)) with C-type antiferromagnetic order~\cite{ding2017unusual,yi2015high,hasan2022perovskite,k47t-23gp}.
For collinear spins, TlCrO$_3$ is altermagnetic, with MSG $Pnma$ (62.441) and MSG without SOC $Pn'ma'$ (62.448). No spin splitting is seen in Fig.~\ref{fig:wide_figureS3}(a).
In Ref.~\onlinecite{k47t-23gp}, altermagnetic spin splitting was shown by adding the lines $\Gamma$-$R$, $U$-$X$, and $X$-$Y$, to the conventional plot.
In contrast, the new band structure plotting approach shows the altermagnetic spin splitting throughout the interior of the BZ, as seen in Fig.~\ref{fig:wide_figureS3}(b).  Computation of the band structure with DFT+U, U value of 4 eV, yields a band gap of 0.52 eV, to be compared with a metallic state with DFT.

Experimental and theoretical studies of HoFeO$_3$ show that it has an orthorhombic distorted perovskite structure (parent space group $Pnma$ (62)) with G-type antiferromagnetic order, and MSG $Pn'ma'$ (62.448)~\cite{chatterji2017temperature,sadhukhan2022multiferroic,khaliq2026spin}.
We find that HoFeO$_3$ has MSG without SOC $Pn'ma'$ (62.448), as presented in Table~\ref{tab:tab1}. 
There is no spin splitting in the conventional band structure plot, presented in Fig.~\ref{fig:wide_figureS4}(a). Our systematic approach for band structure plotting shows spin splitting throughout the interior of the BZ, as seen in Fig.~\ref{fig:wide_figureS4}(b). Our DFT+U computation with Hubbard U value of 5.3 eV on Fe results in a band gap of 0.86 eV, an increase from the DFT value of 0.19 eV, as can be seen in Table~\ref{tab:tab1}.
\begin{figure*}[h] 
    \centering
    \includegraphics[width=0.8\textwidth]{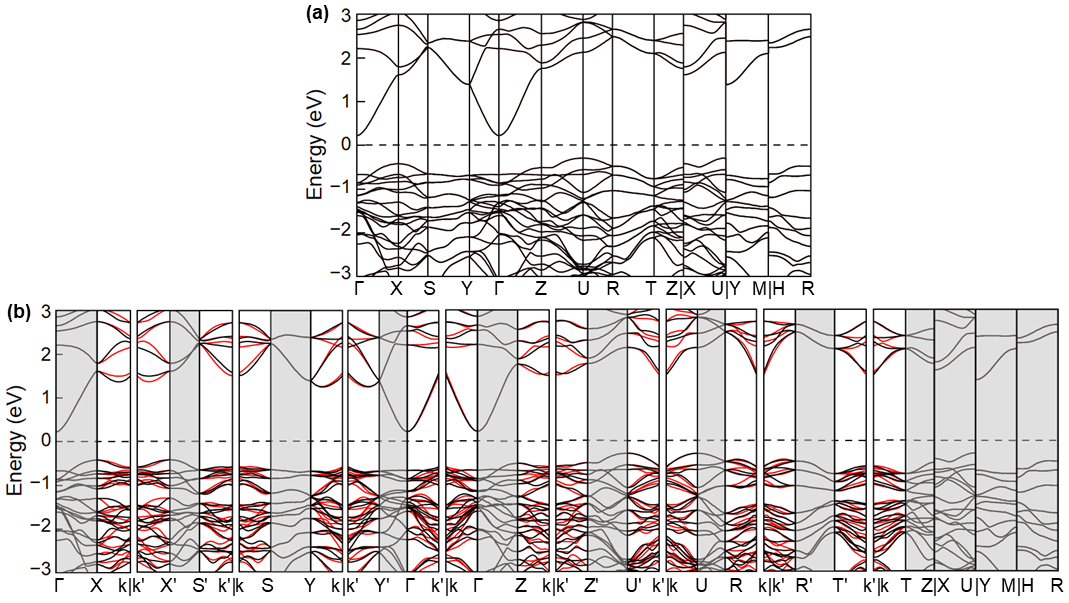} 
    \vspace{-0.5cm}
    \begin{justify}
     \caption{Nonrelativistic DFT+U band structures of orthorhombic HoFeO$_3$ (U = 5.3~eV) (a) The band structure along high-symmetry lines; (b) the band structure along the generalized BZ path constructed using AlterSeeK-Path.}
    \label{fig:wide_figureS4}
    \end{justify}
\end{figure*}

Experimental and theoretical studies of InCrO$_3$ show that it has an orthorhombic distorted perovskite structure (parent space group $Pnma$ (62)) with C-type antiferromagnetic order and MSG $Pnma$ (62.441)~\cite{ding2017unusual,belik2012crystal}. Our analysis shows that it has MSG without SOC $Pn'ma'$ (62.448). 
There is no spin splitting in the conventional band structure plot, presented in Fig.~\ref{fig:wide_figureS5}(a). Our systematic approach for band structure plotting shows spin splitting throughout the interior of the BZ, as seen in Fig.~\ref{fig:wide_figureS5}(b). Our DFT+U computation with Hubbard U value of 4 eV on Cr results in a band gap of 1.73 eV, an increase from the DFT value of 0.94 eV, as can be seen in Table~\ref{tab:tab1}.

Finally, we considered DyFeO$_3$, which is reported in experimental and theoretical studies to have an orthorhombic distorted perovskite structure (parent space group $Pnma$ (62)) with G-type antiferromagnetic order and MSG $Pnma$ (62.441)~\cite{ritter2022magnetic,tokunaga2008magnetic,wang2016simultaneous,stroppa2010multiferroic}. We find that its MSG without SOC is $Pn'ma'$ (62.448). 
As shown in Fig.~\ref{fig:wide_figureS6}(a), the conventional band structure plot shows zero splitting. Our systematic approach for band structure plotting shows spin splitting throughout the interior of the BZ, as seen in Fig.~\ref{fig:wide_figureS6}(b). Band structure computation with DFT+U, U value of 5.3 eV, shows a metallic character. 
\begin{figure*}[h] 
    \centering
    \includegraphics[width=0.8\textwidth]{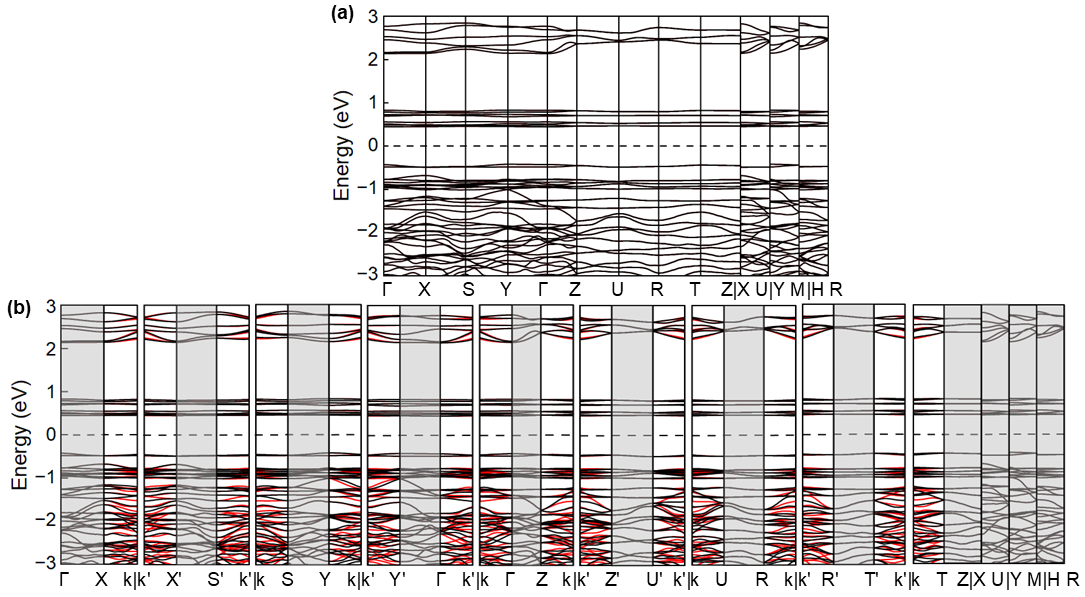} 
    \vspace{-0.5cm}
    \begin{justify}
      \caption{Nonrelativistic DFT+U band structures of orthorhombic InCrO$_3$ (U = 4~eV) (a) The band structure along high-symmetry lines; (b) the band structure along the generalized BZ path constructed using AlterSeeK-Path.}
    \label{fig:wide_figureS5}  
    \end{justify}
\end{figure*}
\begin{figure*}[h] 
    \centering
    \includegraphics[width=0.8\textwidth]{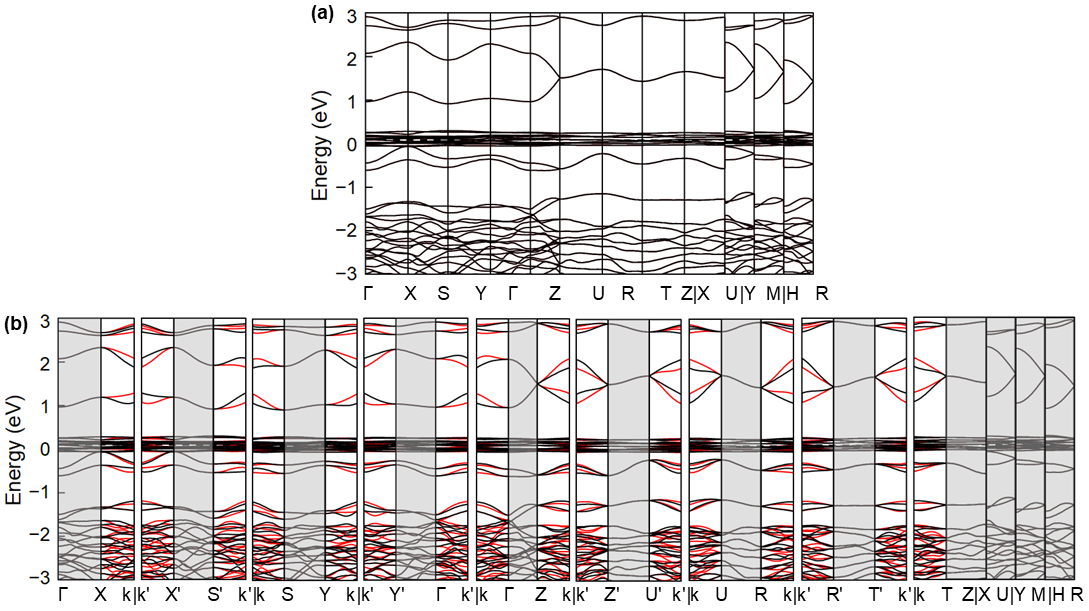} 
    \vspace{-0.5cm}
    \begin{justify}
\caption{Nonrelativistic DFT+U band structures of orthorhombic DyFeO$_3$ (U = 5.3~eV) (a) The band structure along high-symmetry lines; (b) the band structure along the generalized BZ path constructed using AlterSeeK-Path.}
    \label{fig:wide_figureS6}
    \end{justify}
\end{figure*}
\clearpage
\section{DFT+U calculations: choice of U}
In Fig.~\ref{fig:wide_figureS7}(a)-(b), we present the band structures for MnTe along high-symmetry lines with Hubbard U values of 4 eV and 8 eV, respectively.  While the DFT (U = 0~eV) calculation shows a metallic band structure, these DFT+U calculations show that the nonzero U values open the band gap to 0.83, 0.96 and 1.05 eV for U = 4, 6, and 8~eV, respectively, as reported in Table~\ref{tab:tab1}; the U = 6~eV band structure is shown in the main text.
On the other hand, with nonzero Hubbard U values of 6 eV and 8 eV on Cr for CrSb, the DFT+U calculations show a metallic band structure, consistent with experiment (Fig.~\ref{fig:wide_figureS8}(a)-(b) and Table~\ref{tab:tab1}).
\begin{figure*}[h] 
    \centering
    \includegraphics[width=0.8\textwidth]{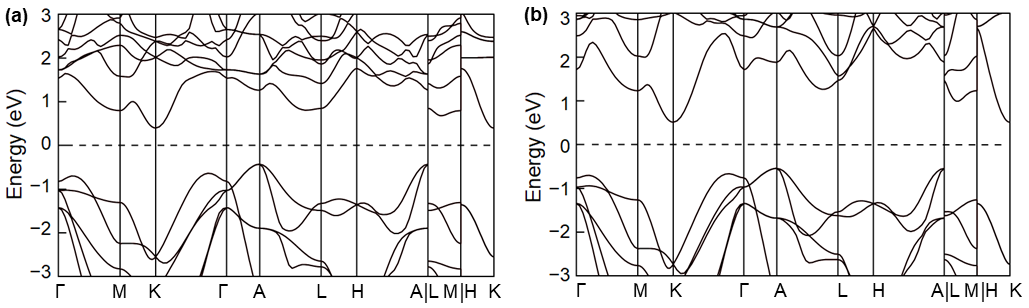} 
    \vspace{-0.5cm}
    \begin{justify}
      \caption{The DFT+U band structures of MnTe along high-symmetry lines with (a) U = 4~eV and (b) U = 8~eV.}
    \label{fig:wide_figureS7}  
    \end{justify}
\end{figure*}
\begin{figure*}[h] 
    \centering
    \includegraphics[width=0.8\textwidth]{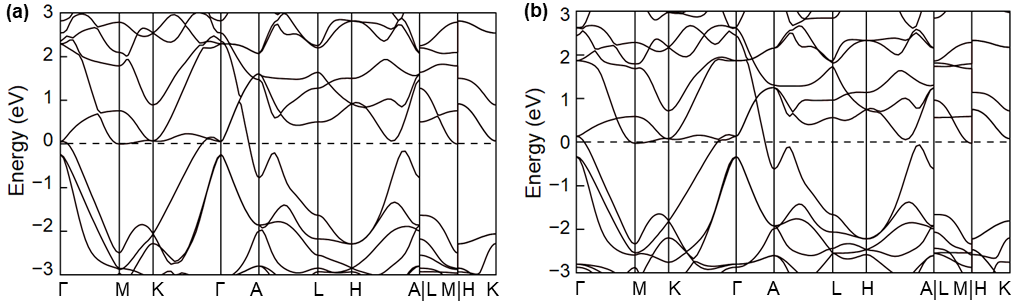} 
    \vspace{-0.5cm}
    \begin{justify}
\caption{The DFT+U band structures of CrSb along high-symmetry lines with (a) U = 6~eV and (b) U = 8~eV.}
    \label{fig:wide_figureS8}
    \end{justify}
\end{figure*}
\end{justify}
\clearpage

\bibliography{cite}